\documentstyle[12pt]{article}

\def\be{\begin{equation}}
\def\ee{\end{equation}}
\def\bea{\begin{eqnarray}}
\def\eea{\end{eqnarray}}
\def\ba{\begin{array}}
\def\ea{\end{array}}
\def\ben{\begin{enumerate}}
\def\een{\end{enumerate}}

\def\nnu{\nonumber}
\def\lll{\label}
 
\begin{document}
\newcommand{\half}{{\textstyle\frac{1}{2}}}
\newcommand{\eqn}[1]{(\ref{#1})}
\newcommand{\npb}[3]{ {\bf Nucl. Phys. B}{#1} ({#2}) {#3}}
\newcommand{\pr}[3]{ {\bf Phys. Rep. }{#1} ({#2}) {#3}}
\newcommand{\prl}[3]{ {\bf Phys. Rev. Lett. }{#1} ({#2}) {#3}}
\newcommand{\plb}[3]{ {\bf Phys. Lett. B}{#1} ({#2}) {#3}}
\newcommand{\prd}[3]{ {\bf Phys. Rev. D}{#1} ({#2}) {#3}}
\newcommand{\hepth}[1]{ [{\bf hep-th}/{#1}]}
\newcommand{\grqc}[1]{ [{\bf gr-qc}/{#1}]}
 
\def\a{\alpha}
\def\b{\beta}
\def\g{\gamma}\def\G{\Gamma}
\def\d{\delta}\def\D{\Delta}
\def\ep{\epsilon}
\def\et{\eta}
\def\z{\zeta}
\def\t{\theta}\def\T{\Theta}
\def\l{\lambda}\def\L{\Lambda}
\def\m{\mu}
\def\f{\phi}\def\F{\Phi}
\def\n{\nu}
\def\p{\psi}\def\P{\Psi}
\def\r{\rho}
\def\s{\sigma}\def\S{\Sigma}
\def\ta{\tau}
\def\x{\chi}
\def\o{\omega}\def\O{\Omega}
\def\k{\kappa}
\def\pa {\partial}
\def\ov{\over}
\def\br{\\}
\def\ud{\underline}
\begin{flushright}
UM-TH-00-16\\
SINP/TNP/00-20\\
hep-th/0007168\\
\end{flushright}
\bigskip\bigskip
\begin{center}
{\large\bf 
$SL(2, Z)$ Duality and 4-Dimensional 
 Noncommutative  Theories}
\vskip1cm
{\sc J. X. Lu$^a$, 
Shibaji Roy$^b$ and
Harvendra Singh$^b$}
\vskip0.5cm
$^a$ Randall Physics Laboratory \\ University of Michigan, Ann Arbor, 
MI 48109-1120, USA
 \vskip 0.5cm 
$^b$ Theory Division, Saha Institute of Nuclear Physics,\\
1/AF Bidhannagar, Calcutta-700 064, India 
\vskip 0.5cm
E-mails: jxlu@umich.edu; roy, hsingh@tnp.saha.ernet.in
\end{center}
\bigskip
\centerline{\bf ABSTRACT}
\bigskip

\begin{quote}
  
We investigate how the four-dimensional noncommutative open 
string/Yang-Mills theory behaves under a general non-perturbative quantum
$SL(2,Z)$ symmetry transformation. We discuss this by considering D3
branes in a constant background of axion, dilaton, and electric and 
magnetic fields (including both $ {\bf E} \perp {\bf B}$ and 
{\bf E}$||${\bf B} cases) in the respective decoupling limit. We find
that the value of axion, whether rational or irrational, determines the 
nature of the resulting theory under $SL(2,Z)$ as well as  its
properties such as the coupling constant and the number of noncommutative 
directions. In particular, a strongly coupled theory with an irrational
value of axion can never be physically equivalent to a weakly coupled
theory while this is usually true for a theory with a rational value of axion.
A noncommutative Yang-Mills (NCYM) (resulting from D3 branes with pure
magnetic flux) is physically equivalent to a noncommutative open string
(NCOS) but if the value of axion is irrational, we also have
noncommutative space-space directions in addition to the usual
noncommutative space-time directions for NCOS. We also find in general 
that a NCOS cannot be physically equivalent to a NCYM but to another NCOS
if the value of axion is irrational. We find another new decoupling
limit for possible light-like NCYM whose $SL(2,Z)$ duality is a
light-like ordinary Yang-Mills if the value of the axion is rational.
Various related questions are also discussed.

\end{quote} 

\newpage
\section{Introduction}
There is a great surge of interest recently on noncommutativity along
space-time directions in string/M theory \cite{seisut,
seist,gopmms,gars,barr,gomm,
gop,klem,berbss,har,chew,lursone,russj,ahagm,kawt,bers,reyv,harone}. This
noncommutativity implies a stringy uncertainty principle which appears
as a special case of a fundamental one in string/M theory as advocated 
in \cite{yo}. 
A field theory defined on such a noncommutative geometry cannot be unitary
\cite{seisut,gomm,kurr} and therefore if there exists a decoupled
theory on such geometry in string/M theory, it cannot be a field theory.  
In the context of D-branes with pure electric flux, it was shown in 
\cite{seist} 
that such a decoupled theory indeed exists and is a noncommutative
open 
string (NCOS) living on the brane. In obtaining such a decoupled theory,
the
electric field must be close to its critical value such that it almost 
balances the original string tension. We therefore end up effectively
with an almost tensionless string. Such a tiny string tension defines a
new energy scale for the decoupled NCOS as well as  the scale for the 
noncommutativity\footnote{We can scale the original tension to infinity while
keeping the new effective tension fixed.}. In this decoupling limit, the
closed string coupling blows up while the coupling for the decoupled
NCOS remains fixed and therefore is small in comparison with the
infinitely large closed string coupling. This also implies that the
transition of NCOS into closed strings cannot occur easily, indicating that
the NCOS decouples from the bulk closed strings. 

    The same conclusion for the existence of NCOS was also obtained in 
\cite{gopmms} but from a rather different viewpoint. Consider a
4-dimensional noncommutative Yang-Mills (NCYM) which is the decoupled 
field theory of D3 branes in a purely magnetic field background. The
decoupling
limit for this theory requires the closed string coupling to scale to
zero. When the gauge coupling for the NCYM is large, the natural way to
deal with this theory is to go to its S-dual description. The authors in
\cite{gopmms} asked:  What is the S-dual of this NCYM? A
worldvolume magnetic field is mapped to an electric field under 
S-duality\footnote{If one instead uses a $B$-field with a nonvanishing
spatial component, one cannot
reach the conclusion directly from the worldvolume point of view since a
spatial $B$-field under S-duality turns into a spatial RR 2-form field.
One often says that D3-branes (in general Dp-branes) 
with spatial nonvanishing $B$-field give rise
to NCYM in the corresponding decoupling limit. From the field theory (or
open string point of view) side, this is correct since all we need for the
NCYM is
the closed string metric, closed string coupling and the $B$-field as 
demonstrated in \cite{seiw}. However, from the gravity description of 
D3-branes with spatial nontrivial $B$-field,  we must also have D-strings
present such that the resulting configuration is BPS. These D-strings
require a nonvanishing space-time component of RR 2-form. Under
S-duality, this nonvanishing component becomes a $B$-field which is needed
for the space-time noncommutativity as demonstrated in
\cite{har,lursone}. So this seems to indicate that the gravity
description of NCYM contains more information than the field theory
description. For example, in order to have the correct decoupling limit
for NCOS, we have to perform the non-linear S-duality on the worldvolume 
while on the gravity side the S-duality is linear. This is the precise
reason that we chose to work on a concrete gravity system of ((F, D1),
D3) bound state for the decoupled NCOS with noncommutativity in both
space-time and space-space directions in \cite{lursone}. But in the prsent 
paper, we choose to work in 
the hard way, i.e. from the worldvolume side.}. 
Moreover, it was found that the decoupling limit in the S-dual theory
exists only when the electric field attains a critical value. 
Also, in this case the original vanishing closed string coupling is
transformed to its inverse in the S-dual, therefore becoming
infinitely large. 
 So, according to what has been obtained in \cite{seist}, the S-dual
of NCYM is a NCOS.

	The NCYM used in \cite{gopmms} arises as a decoupled theory 
of D3 branes with a pure magnetic flux. We can have  
decoupled NCYM  from D3 branes with both electric and magnetic
fields as discussed in \cite{chew,lursone,russj} and further with
nonvanishing axion which will be discussed in this paper\footnote{While
we were writing up this paper, we became aware a paper \cite{cai} in
which the $SL(2,Z)$ duality of the gravity dual description of ((F, D1),
D3) bound state, corresponding to ${\bf E}||{\bf B}$ case considered
here, appeared on the net.}. We can also
have decoupled NCOS from D3 branes in the presence of both 
electric and magnetic fields as discussed in \cite{chew,lursone,russj,cai}
and further with a nonvanishing axion which, for the ${\bf E}||{\bf B}$
case, has been discussed in \cite{russj,cai}.  
One may wonder if the S-dual or in general a $SL(2,Z)$ dual of NCYM 
(or NCOS) always give  a NCOS (or NCYM). In \cite{lursone}, the
present authors showed
that the S-dual of NCYM does not in general give a NCOS. In 
\cite{russj} for the case {\bf E}$ ||${\bf B}, it was shown that a NCOS is
mapped
to another NCOS under a $SL(2,Z)$ tranformation if the value of axion 
is irrational. Only for rational value of axion, a NCOS is physically 
equivalent to a NCYM.

      We here intend to give a systematic study of the $SL(2,Z)$ dual
of NCYM/NCOS which arises from D3 branes with nonvanishing axion and
in the presence of both electric and magnetic fields in the respective
decoupling limits. We will consider both of {\bf E}$ ||${\bf B} and {\bf 
E}$\perp${\bf B} cases.
Our plan is as follows: In the following section, we
give explicit $SL(2,Z)$ transformation rules for the worldvolume gauge
fields and the generalized worldvolume gauge fields in the most general
background. In section 3, we will calculate quantities relevant for 
decoupling limits for NCYM/NCOS in two versions related by $SL(2,Z)$
duality using Seiberg-Witten relations\cite{seiw}. We consider both
{\bf E}$\perp${\bf B} and {\bf E}$||${\bf B} cases. 
In section 4, we will discuss various decoupling limits for NCOS/NCYM
and the
$SL(2,Z)$ dual of NCYM/NCOS. We conclude this paper in section 5.

\section{D3 Branes and Nonperturbative $SL(2,Z)$ Symmetry}

 Type IIB string theory is conjectured to have a nonperturbative
quantum $SL(2,Z)$ symmetry. This symmetry implies that two Type IIB string 
theories related by an $SL(2,Z)$ transformation are physically equivalent.
Among various dynamical objects in this theory,  D3 branes play a
special role in the sense that this object, like type IIB string
itself, is invariant under the $SL(2,Z)$. For the purpose of this paper,
we need to consider only the bosonic sector of the low energy effective 
action for a single D3 brane coupled to a most general background.
This is described by a Born-Infeld type action \cite{tse,greg,cedgnw} in
Einstein-frame as
\be
S_4 = - \frac{1}{(2 \pi)^3 \a'^2} \int d^4 x \sqrt{- \det
(g^E + e^{-\phi/2} {\cal F})} + \frac{1}{(2 \pi)^3
\a'^2} \int d^4 x (C \wedge e^{\cal F})_4,
\ee  
where ${\cal F}_2 = 2 \pi \a' F_2 + B_2$ and
\be
(C \wedge e^{\cal F})_4 = C_4 + C_2 \wedge {\cal F} + \frac{1}{2} C_0
{\cal F}_2 \wedge {\cal F}_2.
\ee
In the above, the 2-form $F_2$ is the worldvolume $U(1)$ field strength,
the worldvolume metric is the pullback of spacetime metric,
$B_2$ is the pullback of spacetime NSNS 2-form potential and $C_n$ is the 
pullback of spacetime RR $n$-form potential. In the following, we take
$\mu,\nu = 0, 1, 2, 3$ as worldvolume indices and we will
denote $C_0 = \chi$ for notational convenience.

       For vanishing $C_2$ and $B_2$, it was shown in \cite{gibr} that the
       equations of motion from the above action are just special
forms of a more general action which possess a classical $SL(2,R)$
symmetry provided the dilaton and the axion parametrize the coset 
$SL(2,R)/SO(2)$. It was further shown in \cite{tse,greg} that with the 
inclusion of both $B_2$ and $C_2$, the equations of motion still have
such $SL(2,R)$ symmetry provided the metric, dilaton, axion, 
       $B_2$, $C_2$ and $C_4$ transform according to the rules
       determined from the type IIB supergravity, i.e.,
\bea
&& g^E_{\mu\nu} \to g^E_{\mu\nu},~~ C_4 \to C_4,~~ \lambda \to 
\frac{a \lambda + b}{c \lambda + d},\nnu\\
&& \left(\begin{array}{c} B_2\\C_2\end{array}\right) \to 
\left(\Lambda^{-1}\right)^T \left(\begin{array}{c} B_2\\C_2\end{array}\right), 
\lll{i3}\eea
where the complex scalar\footnote{Alternatively, the dilaton and axion
can be used to parameterize the coset $SL(2,R)/SO(2)$ as
$$ {\cal M} = \left(\begin{array}{cc}
                     \chi^2 + e^{-2\phi}&\chi\\
                      \chi&1\end{array}\right) e^\phi. $$
Then under $SL(2, R)$, ${\cal M} \to \Lambda {\cal M} \Lambda^T$ with
$SL(2, R)$ matrix $\Lambda$ defined in \eqn{i4}.}
  $\lambda = \chi + i e^{-\phi}$,
  $\Lambda$ is a $2\times2$  $SL(2,R)$ matrix defined as
\be
\Lambda = \left(\begin{array}{cc}a&b\\c&d \end{array}\right), ~~~ad - bc = 1,
\lll{i4}\ee 
and `T' denotes the transpose of the matrix.

    Our goal in this section is to express the transformed $F_2$ (or
    ${\cal F}_2$) under $SL(2,R)$ in terms of the original fields 
in a simple way. As is understood, the $SL(2,R)$ symmetry is manifest
    only on equations of motion and it rotates between equations of
    motion and Bianchi identities for $F_2$. Let us define a quantity
$K^{\mu\nu}$ as 
\be
\sqrt{-\det g^E} \frac{K^{\mu\nu}}{2\pi} = \frac{\delta
    S_4}{\delta F_{\mu\nu}}. \lll{i5}
\ee
Note that
\be
 \det \left(g^E + e^{-\phi/2} {\cal F}\right) = \left(\det g^E \right)
\left(1 + \frac{1}{2} e^{-\phi} {\cal F}^2 - \frac{1}{16}
    e^{-2\phi}\left({\cal F} \star {\cal F}\right)^2 \right),\lll{i6}
\ee
where $\star$ denotes the Hodge-dual on the brane. With the above, we
have
\be
2 \pi \a' K_{\mu\nu} = - \frac{e^{-\phi} {\cal F}_{\mu\nu} - \frac{1}{4} 
e^{- 2\phi} \left({\cal F} \star {\cal F}\right) (\star {\cal
    F})_{\mu\nu}} {\sqrt{1 + \frac{1}{2} e^{-\phi} {\cal F}^2 -
    \frac{1}{16} e^{- 2 \phi} \left({\cal F} \star{\cal F}\right)^2}}
+ (\star C)_{\mu\nu} + \chi (\star{\cal F})_{\mu\nu}.\lll{i7}
\ee
With the above expression, the equation of motion for gauge potential $A$ 
($F_2 = d A$) is
\be
d \star K_2 = 0. \lll{i8}
\ee
So combining with Bianchi identity $d F_2 = 0$, we have
\be
d \left(\begin{array}{c} F_2\\\star K_2 \end{array}\right) = 0.\lll{i9}
\ee
Given any solution of the above equation for $F_2$ ($K_2$ is given
through \eqn{i7}), it appears that we could obtain another solution
from this through a global $GL (2, R)$ rotation. But, since D3-branes 
 appear as sources to the bulk
gravity, the energy-momentum tensor due to this source must be kept
invariant under this symmetry since the Einstein-frame metric is inert
to this symmetry. Further the equations of motion
 for various potentials in the bulk spacetime should be transformed 
covariantly under this transformation when the D3 brane source is
considered. The global symmetry for the bulk gravity therefore restricts 
us
to have only a global classical $SL(2,R)$ rather than $GL (2,R)$ for the
D3 brane. With the D3 brane as source, we can deduce from equations of
motion for  $B_2$ and $C_2$,  that $(F_2, \star K_2)$ transform in the
same way as $(B_2, C_2)$ under $SL(2,R)$, i.e.,
\be
\left(\begin{array}{c} F_2\\\star K_2 \end{array}\right) \to
\left(\Lambda^{-1}\right)^T  
\left(\begin{array}{c} F_2\\\star K_2 \end{array}\right).      
\lll{i10} \ee
We can define a generalized 2-form ${\cal K}_2$ as
\be
{\cal K}_2 = 2\pi \a' \star K_2 + C_2,
\lll{i11} \ee
analogous to ${\cal F} = 2\pi \a' F_2 + B_2$. Given the transformations
for $(F_2, \star K_2)$ and $(B_2, C_2)$ under $SL(2, R)$, we have
\be
\left(\begin{array}{c} {\cal F}_2\\ {\cal K}_2 \end{array}\right) \to
\left(\Lambda^{-1}\right)^T  
\left(\begin{array}{c} {\cal F}_2\\ {\cal K}_2 \end{array}\right).
\lll{i12} \ee
This is the key equation which we will use in the following section.
With $K_2$ given in \eqn{i7}, \eqn{i11} can be re-expressed as
\be
- {\cal K}_2 = \frac{e^{-\phi} (\star{\cal F})_2 + \frac{1}{4} e^{-2\phi}
({\cal F}\star{\cal F}) {\cal F}_2}{\sqrt{1 + \frac{1}{2} e^{-\phi} {\cal F}^2 -
    \frac{1}{16} e^{- 2 \phi} \left({\cal F} \star{\cal F}\right)^2}} +
    \chi {\cal F}_2. \lll{i13}
\ee

The above equation implies the following constraint 
which generalizes the one given in \cite{gibr} as
\be 
({\cal F}_2, {\cal K}_2) {\cal M} 
\left(\begin{array}{c} \star{\cal F}_2\\ \star {\cal K}_2
\end{array}\right) = 0,\lll{i14}
\ee
which is useful for proving the invariance of the energy-momentum
tensor.

In the following section, we will use the above equations \eqn{i12} and
\eqn{i13}
for calculating the open string metric and noncommutativity parameters
from the Seiberg-Witten relations for both ${\bf E}||{\bf B}$ and 
${\bf E}\perp{\bf B}$ cases.

\section{Seiberg-Witten Setup}

	In this section, we calculate the effective open string metric and
noncommutativity parameters for both {\bf E}$\perp${\bf B} and 
 {\bf E}$||${\bf B} cases
using Seiberg-Witten formulae\cite{seiw}. The effective open string 
metric is 
\be 
G_{\m\n}=g_{\m\n}- ({\cal F}~g^{-1}~{\cal F})_{\m\n},
\lll{15}
\ee
and the anti-symmetric noncommutativity parameter is 
\be
\Theta^{\m\n}= 2\pi\a' \left( {1\ov g+ {\cal F}}\right)^{\m\n}_A,
\lll{16}\ee
where `A' denotes the antisymmetric part. The effective open string
coupling $G_s$
is related to the closed string coupling $g_s = e^\phi$ through the
following relation:
\be
G_s=g_s \left( { det G\over
det(g+ {\cal F})}\right)^{1/2}
\lll{17}
\ee

As usual, we assume  in the following that
${\cal F}_2$ is entirely given by the worldvolume 
field $F_2$ and set the NSNS $B_2$ to zero. In other 
words, we trade NSNS $B_2$ for the worldvolume $F_2$ through a gauge 
transformation.  
For either {\bf E}$\perp${\bf B} or {\bf E} $||${\bf  B} case, 
we calculate the above open
string quantities from the relevant closed quantities and
the worldvolume ${\cal F}$ in two versions related
by $SL(2,Z)$-duality. From now on, we limit ourselves to the
non-perturbative quantum $SL(2, Z)$ symmetry rather than the classical
$SL(2,R)$. In other words, we consider physically equivalent theories
related by $SL(2,Z)$. For convenience, let us write down the transformed
$e^{\hat \phi}$, $\hat \chi$, $\hat {\cal F}_2$ and $\hat {\cal K}_2$ 
in terms of the corresponding original fields and  
integral $SL(2,Z)$ elements $a, b, c,
d$ which satisfy $ad - bc = 1$ as
\bea
&&e^{\hat\phi} = e^\phi |c \lambda + d|^2, ~~~
\hat\chi = \frac{ac(\chi^2 + e^{- 2\phi}) + (ad + bc)\chi + bd}{|c
\lambda
  + d|^2},\nnu\\
&&\hat {\cal F}_2 = d {\cal F}_2 - c {\cal K}_2, ~~ \hat {\cal K}_2 = - b {\cal
F}_2 + a {\cal K}_2. \lll{19}
\eea
where ${\cal K}_2$ is given by \eqn{i13}. The string metric is defined as 
$g_{\mu\nu} = e^{\phi/2} g^E_{\mu\nu}$ and so we have
\be
\hat g_{\mu\nu} = g_{\mu\nu} |c \lambda + d|.
\lll{20}\ee 
We always denote the corresponding quantities in the $SL(2,Z)$ dual 
with `hat' over the letters as indicated above.  
 Let us begin with the  {\bf E}$\perp${\bf B} case first.

\subsection{E$\perp$B Case}

	Our starting point is to choose constant $F_{\mu\nu}$ 
\be
{\cal F}_2 = 2\pi \a' \left(\begin{array}{cccc}
                             0&E&0&0\\
                             - E&0&B&0\\
                             0&- B&0&0\\
                             0&0&0&0 \end{array}\right),
\lll{21}\ee
and the
constant closed string metric in string-frame as 
$g_{\mu\nu} = {\rm diag}(- g_0, g_1, g_2, g_3)$ with $g_0, g_1, g_2,
g_3$ all positive fixed parameters. 

 Using \eqn{15}, we have the open string metric as
\be
G_{\mu\nu} = \left(\begin{array}{cccc}
                    - g_0 (1 - {\tilde E}^2) & 0 & - \sqrt{g_0 g_2}
		    \tilde E \tilde B&0\\
                   0& g_1 (1 - {\tilde E}^2 + {\tilde B}^2)&0&0\\
                   -\sqrt{g_0 g_2} \tilde E\tilde B&0&g_2 (1 + {\tilde
		   B}^2)&0\\
                   0&0&0&g_3\end{array}\right),
\lll{22}\ee
and using \eqn{16} the noncommutativity parameter as
\be
\Theta^{\mu\nu} = \frac{1}{1 - {\tilde E}^2 + {\tilde B}^2}
                 \left(\begin{array}{cccc}
                       0& \tilde E / E_0&0& 0\\
                       - \tilde E /E_0&0 & - \tilde B /B_0 &0\\
                       0& \tilde B /B_0 & 0 &0\\
                       0 & 0 & 0 & 0\end{array} \right).
\lll{23}\ee
The open string coupling can be obtained from eq.\eqn{17} as,
\be
G_s = g_s \left(1 - {\tilde E}^2 + {\tilde B}^2\right)^{1/2},
\lll{24}\ee
which implies that the critical field in this case is 
$(1 + \tilde B^2)^{1/2}$.
In the above, we have defined
\be
\tilde E = \frac{E}{E_0},~~ \tilde B = \frac{B}{B_0},
\ee
where the parameters $E_0 = \sqrt{g_0 g_1}/(2\pi \a')$ and
$B_0 = \sqrt{g_1 g_2} / (2 \pi \a')$.                         

     One might think that the {\bf E}$\perp${\bf B} case is simpler 
than the {\bf E}$||${\bf  B}
case. On the contrary, it is a bit more complicated in both the
decoupling limits (NCYM and NCOS) which will be studied in the following
section 
and the $SL(2,Z)$ dual formulation. Let us derive the open string
metric, the noncommutativity parameter and the open string coupling in the
$SL(2, Z)$ dual. In order to calculate these quantities, we 
have to express $\hat {\cal F}_2$ in terms of relevant quanties
in the
original version. For constant $F_{\mu\nu}$, the equation of motion
is satisfied and so we can use the duality relation to calculate
$\hat{\cal F}_2$. In doing so, we first need to calculate $ {\cal K}_2$
from \eqn{i13}. Thus we find,
\be
{\cal K}_{\mu\nu} = - \left(\begin{array}{cccc}
  0&\sqrt{g_0 g_1} \tilde E \chi&0& - \sqrt{g_0 g_3} \tilde B/G_s\\
  - \sqrt{g_0 g_1} \tilde E \chi&0 & \sqrt{g_1 g_2} \tilde B \chi& 0\\  
  0& - \sqrt{g_1 g_2} \tilde B \chi &0 & \sqrt{g_2 g_3} \tilde E/G_s\\
  \sqrt{g_0 g_3} \tilde B /G_s& 0& - \sqrt{g_2 g_3} \tilde E /G_s& 0
\end{array}\right),
\lll{26}\ee
where we have used $g_{\mu\nu} = g_s^{1/2} g^E_{\mu\nu}$ and the
relation between $G_s$ and $g_s$ given in \eqn{24} as well as the
definitions for $\tilde E$ and $\tilde B$ given above.
With the above and $\hat {\cal F}_2 = d~ {\cal F}_2 - c~ {\cal K}_2$, we 
have
\be
\hat{\cal F}_{\mu\nu} = \left(\begin{array}{cccc}
  0&\sqrt{g_0 g_1} \tilde E (c\chi + d)&0& - \sqrt{g_0 g_3} c \tilde B/G_s\\
  - \sqrt{g_0 g_1} \tilde E (c \chi + d)&0 & \sqrt{g_1 g_2} \tilde B (c
  \chi + d)& 0\\  
  0& - \sqrt{g_1 g_2} \tilde B (c \chi + d) &0 & \sqrt{g_2 g_3} c \tilde E/G_s\\
  \sqrt{g_0 g_3} c \tilde B /G_s& 0& - \sqrt{g_2 g_3} c \tilde E /G_s& 0
\end{array}\right)
\lll{27}\ee
We notice from the expression of $\hat{\cal F}_{\m\n}$ in \eqn{27}
 that we now have additional electric and
magnetic fields pointing along negative $x^3$-direction and
$x^1$-direction, respectively, in the $SL(2,Z)$ dual even though we 
originally had only
electric field pointing along $x^1$-direction and magnetic field
pointing along $x^3$-direction. This implies that we may have more
noncommutative directions in the $SL(2,Z)$ dual. This differs from
{\bf E}$||${\bf B} case as we will see.

Using again Seiberg-Witten relations and after some tedious calculations,
we find the new open string metric to have the form 
\be
\hat{G}_{\mu\nu} = \frac{|c S + d|^2}{|c \lambda + d|} G_{\mu\nu},
\lll{28}\ee
and the noncommutativity parameters take the form, 
\bea
&&\hat{\Theta}^{01} = \frac{(c \chi +d)}{|c S + d|^2}
\Theta^{01},~~
 \hat{\Theta}^{03} = - \frac{2 \pi \a'}{\sqrt{g_0 g_3}} \frac{c \tilde
B / G_s}{|c S + d|^2},\nnu\\
&&\hat{\Theta}^{12} = \frac{(c \chi + d)}{|c S + d|^2}
\Theta^{12},~~ \hat{\Theta}^{23} = - \frac{2 \pi \a'}{\sqrt{g_2 g_3}} 
\frac{c \tilde E/G_s} {|c S + d|^2},
\lll{29}\eea
where $\lambda = \chi + i/g_s$ defined before, 
$S  = \chi + i/G_S$ 
(since ${\bf E}\cdot {\bf B} = 0$ here), and $G_{\mu\nu}$, $\Theta^{01}$
and $\Theta^{12}$ are the original open string metric, noncommutative 
parameters given in \eqn{22} and \eqn{23}, respectively.
The open string coupling here is related to the original open string
coupling as
\be
\hat{G}_s = |c S + d|^2 G_s.\lll{30}
\ee

\subsection{E $||$ B Case}

	      We now take 

\be
{\cal F} = 2\pi \a' \left(\begin{array}{cccc}
0&E&0&0\\
- E&0&0&0\\
0&0&0&B\\
0&0&- B&0\end{array}\right),
\lll{31}\ee
and the string frame metric $g_{\mu\nu} = {\rm diag}(- g_1, g_1, g_2,
g_2)$ where $g_1,~g_2$ are positive  parameters.
Using Seiberg-Witten relations, we have the open string metric as
\be
G_{\mu\nu} = \left(\begin{array}{cccc}
- g_1 (1 - \tilde{E}^2)&0&0&0\\
0&g_1(1 - \tilde{E}^2)&0&0\\
0&0&g_2 (1 + \tilde{B}^2)&0\\
0&0&0&g_2 (1 + \tilde{B}^2)\end{array}\right),
\lll{32}\ee
the noncommutativity parameters as,
\be
\Theta^{01} = \frac{\tilde E}{E_{c} (1 - \tilde{E}^2)},~~\Theta^{23} =
- \frac{\tilde B}{B_0 (1 + \tilde{B}^2)},
\lll{33}\ee
and the open string coupling as
\be
G_s = g_s (1 - \tilde{E}^2)^{1/2} (1 + \tilde{B}^2)^{1/2}.
\lll{34}\ee
In the above, we have defined $\tilde E = E /E_{c}$ and $\tilde B = B
/B_0$ with the critical field $E_{c} = g_1/(2\pi\a')$ and $B_0 =
g_2/(2\pi \a')$.

We now calculate the relevant open string quantities in the $SL(2,Z)$
 dual. To do so,  we
need to have $\hat {\cal F}_2$ and as before we calculate ${\cal
K}_{\mu\nu}$ first.
Using \eqn{i13}, we have
\be
{\cal K}_{01} = - g_1\left[\tilde E \chi - 
\tilde B (1 - \tilde{E}^2)/G_s\right],
~~
{\cal K}_{23} = - g_2 \left[\tilde B \chi + 
\tilde E (1 + \tilde{B}^2)/G_s)\right],
\lll{35}\ee
where we have used the relation between $G_s$ and $g_s$ given in \eqn{34}, 
 the definitions for $\tilde E$ and $\tilde B$ given earlier and
 $g_{\mu\nu} = g_s^{1/2} g^E_{\mu\nu}$. 
With the above, we have using eq.\eqn{19}
\be
\hat{\cal F}_{01} = g_1\left[\tilde E (c \chi + d) - 
c \tilde B (1 - \tilde{E}^2)/G_s\right],~~
\hat{\cal F}_{23} = g_2 \left[\tilde B (c \chi + d) + 
c \tilde E (1 + \tilde{B}^2)/G_s)\right].
\lll{36}\ee
Using Seiberg-Witten relations \eqn{15}-\eqn{17}, we now have the open
string
 metric
\be
\hat{G}_{\mu\nu} = \frac{|c S + d|^2}{|c \lambda + d|} G_{\mu\nu},
\lll{37}\ee
the noncommutativity parameters
\be 
\hat \Theta^{01} = \frac{(c \chi + d) \Theta^{01} - \frac{c\tilde
 B}{E_{cr} G_s}}{|c S + d|^2}, ~~ \hat \Theta^{23} = \frac{(c \chi + d) 
\Theta^{23} - \frac{c \tilde E}{B_0 G_s}}{|c S + d|^2},
\lll{38}\ee
and the open string coupling
\be
\hat G_s = |c S + d|^2 G_s.
\lll{39}\ee
In the above, $G_s$, $\Theta^{\mu\nu}$ and $G_{\mu\nu}$ are the original
open string coupling, noncommutativity parameters and open string metric, 
respectively. We have now $S = \chi + \tilde E \tilde B/G_s + i/G_s$ for
 which ${\bf E}$ and ${\bf B}$ contribute since ${\bf E}\cdot {\bf B}
 \neq 0$.

\section{Decoupling Limits and $SL(2,Z)$ Duality}

We are now ready to discuss the decouping limits for NCYM/NCOS and the
$SL(2,Z)$ duality for the underlying decoupled theory. Before we discuss
the noncommutative theory, we would like to address one question: Can an 
ordinary theory become a noncommutative theory through $SL(2,Z)$
duality? Our examination gives negative answer. Let us point out that
there is a general
rule regarding whether we can map a strongly coupled theory to a weakly
coupled one through a $SL(2,Z)$ transformation or not for any theory
either ordianry or noncommutative. The rule is: for rational $\chi$, we
can 
have two physically equivalent theories which are strong-weak dual to 
each other while for irrational $\chi$, we do not have this.
We first discuss ${\bf E} \perp {\bf B}$ case and then ${\bf E} ||
{\bf B}$ case. 

\subsection{${\bf E} \perp{\bf B}$ Case}

Let us begin with the decoupling limit for NCYM. To have a NCYM,
we need to decouple not only  the open string ending on the brane
from the closed strings in the bulk but also the open string
massive modes from the massless ones. So we need to send $\a'\to 0$.
To have a sensible quantum theory, we need to fix the open string
coupling and the open string metric in this limit. We also need to fix
at least one nonvanishing spatial component of  the noncommutative
matrix. With these requirements and examining \eqn{22} and \eqn{23}, we
can naively have the
following three limits:

\begin{itemize}
\item 1) \be
\a' \to 0, ~~g_1 = \left(\frac{\tilde b}{\a'}\right)^2,~~\tilde{E}^2 =
1 + \tilde{B}^2 - \left(\frac{\a'}{\tilde b}\right)^2,~~g_s = G_s 
\frac{\tilde b}{\a'}
\lll{40}\ee
with $g_0, g_2, g_3$ and $\tilde B$ fixed. For simplicity, we choose
$g_0 \tilde B^2 = 1$, $g_2 ( 1 + \tilde B^2) = 1$ and $g_3 = 1$. So we have
the metric
\be
G_{\mu\nu} = \left(\begin{array}{cccc}
1 - g_0 (\a'/\tilde b)^2 &0& 1 - g_2 (\a'/\tilde b)^2 /2&0\\
0&1&0&0\\
 1 - g_2 (\a'/\tilde b)^2 /2 &0&1&0\\
0&0&0&1\end{array}\right),
\lll{41}\ee
 and the noncommutativity parameters\footnote{We choose $\tilde B$ to be
 negative for definiteness. For positive $\tilde B$, the discussion and
the conclusion are basically the same.} 
\be
\Theta^{01} =  \Theta^{12} =  2\pi \tilde b |\tilde B| \sqrt{1 + {\tilde
B}^2}.
\lll{42}\ee 

\item 2) 
\be
\a' \to 0,~~ \tilde E = - \tilde B = \left(\frac{\tilde b}{\a'}
\right)^{1/2},~~
g_0 = g_2 = \frac{\a'}{\tilde b}, 
\lll{43}\ee
with $g_1, g_3$ and $g_s$ fixed. We here choose $g_0/g_2 = 1$ just for 
simplicity but in general we need only $g_0/g_2$ fixed. For simplicity,
we also choose $g_1 = g_3 = 1$. Now the open string metric is 
\be
G_{\mu\nu} = \left(\begin{array}{cccc}
1 - \frac{\a'}{\tilde b}&0&1&0\\
0&1&0&0\\
1&0&1 + \frac{\a'}{\tilde b}&0\\
0&0&0&1\end{array}\right),
\ee
and  the noncommutativity parameters are similar to those in 1) as
\be
\Theta^{01} =  \Theta^{12} = 2\pi \tilde b.
\lll{44}\ee
So 1) and 2) are quite similar except for the closed string coupling
$g_s$. $g_s$ blows up in the decoupling limit in 1) while it remains fixed
in 2). This case corresponds to the light-like NCYM discussed in \cite{ahagm}.
\item 3)
\be
\a' \to 0,~~ g_1 = g_2 = \left(\frac{\a'}{\tilde b}\right)^2,~~\tilde B
= \frac{\tilde b}{\a'},~~g_s = G_s \frac{\a'}{\tilde b},
\lll{45}\ee
with $g_0, g_3$ and $\tilde E$ fixed. Here we choose $g_1/g_2 = 1$ for 
simplicity but in general we only need this ratio to be fixed. For the 
present case, we do not have the $\tilde E \le 1$ requirement. It can be
any fixed number or approaching zero. 
 For simplicity we set $g_3 = 1$. This same decoupling
limit for $\tilde E \le 1$ was
discussed in \cite{chew}. With the above, we have the open string metric
\be
G_{\mu\nu} = \left(\begin{array}{cccc}
- g_0 (1 - \tilde E^2)& 0 & - \sqrt{g_0} \tilde E&0\\
0&1&0&0\\
-\sqrt{g_0}\tilde E&0&1&0\\
0&0&0&1\end{array}\right),
\lll{46}\ee
and the nonvanishing noncommutativity parameter
\be
\Theta^{12} = - 2\pi \tilde b.
\lll{47}\ee
\end{itemize}

We  point out that there are both space-time and space-space
noncommutativities in 1) and 2) while there is only space-space
noncommutativity in 3). The space-time noncommutativity arises because
the  electric field  approaches the critical value in both cases.
In general, one expects that the underlying
decoupled theories are NCOS rather than a NCYM.  At least for 2), the 
unitarity discussion given in \cite{ahagm} seems to indicate that the 
resulting theory is a light-like NCYM. Actually, 1) has the same
structure as in 2). So we expect that we might also have a light-like
NCYM in the limit of 1).
The arguments given in \cite{ahagm} are: even though
we have both $\Theta^{01}$ and $\Theta^{12}$ in appearance, if we choose
light-like coordinates $x^\pm = (x^0 \pm x^2)/\sqrt{2}$, work in the 
$(x^+, x^1, x^-, x^3)$-system and take $x^+$ as the light-cone time, we
then have only nonvanishing noncommutativity parameter $\Theta^{-1}$, a 
space-space noncommutativity. In addition, the underlying theory is 
unitary, i.e., the inner product $p \circ p = - p_\mu \Theta^{\mu\rho}    
G_{\rho\sigma} \Theta^{\sigma\nu} p_\nu$ is never negative. Therefore,
it appears that the underlying theory is a well-defined NCYM. Let us 
demonstrate this for both 1) and 2). We now express everything in the 
light-like coordinate $(x^+, x^1, x^-, x^3)$-system. Let us denote the 
corresponding cases as $1^{(lc)})$ and $2^{(lc)})$, respectively.

\begin{itemize}
\item $1^{(lc)}$): We now have
the metric
\be
G^{(lc)}_{\mu\nu} = \left(\begin{array}{cccc}
2&0&0&0\\
0&1&0&0\\
0&0&-  \frac{(\a'/\tilde b)^2}{2 g_0 g_2}&0\\
0&0&0&1\end{array}\right)
\ee
and the only nonvanishing noncommutativity parameter 
\be
\Theta^{(lc) - 1} = - \Theta^{(lc)1-} = 2\sqrt{2} \pi \tilde b |\tilde B| \sqrt{1 + \tilde
B^2}.
\ee
One can check that indeed $p \circ p$ is non-negative as
$p \circ p = - p_\mu \Theta^{\mu\rho} G_{\rho \sigma} \Theta^{\sigma\nu}
p_\nu = (p_{-})^2 (\Theta^{- 1})^2 \ge 0$ when $\a' \to 0$ is taken.
 
\item $2^{(lc)})$: We have the open string metric 
\be
G^{(lc)}_{\mu\nu} = \left(\begin{array}{cccc}
2 & 0 & - \frac{\a'}{\tilde b}&0\\
0&1&0&0\\
- \frac{\a'}{\tilde b}&0&0&0\\
0&0&0&1\end{array}\right),
\ee
and the nonvanishing noncommutativity parameter
\be
\Theta^{(lc)-1} = - \Theta^{(lc)1-} = 2\sqrt{2} \pi \tilde b.
\ee
One can check again $p \circ p = (p_{-})^2 (\Theta^{(lc)-1})^2 \ge 0$.
\end{itemize}

The above discussion seems to indicate that in terms of the light-like
coordinates, $1^{(lc)})$ and $2^{(lc)})$ cases look no different from the
case 3) above. According to the criterion given in \cite{gomm,ahagm},
each of the field theories is unitary and each has a space-space 
noncommutativity. We therefore should call the decoupled field theories in
$1^{(lc)}$) and $2^{(lc)}$) as light-like NCYM.

Let us discuss the $SL(2,Z)$ duality for each of the NCYM. Let us denote
the corresponding cases as $1^{(lc)})'$, $2^{(lc)})'$ and $3)'$,
respectively. 
\begin{itemize}
\item $1^{(lc)})'$ We need to consider irrational $\chi$ and rational
$\chi$ separately. a) If $\chi$ is irrational, $|c \lambda + d| = 
(c\chi + d)$ since $g_s \to \infty$ in the decoupling limit 1).
$|c S + d|^2 = (c\chi + d)^2 + c^2/G_s^2$ remains fixed. Using \eqn{28},
\eqn{29}, \eqn{30} and the decoupling limit in 1), we have
\be
\hat G^{(lc)}_{\mu\nu} = \frac{|c S + d|^2}{c \chi + d}
G^{(lc)}_{\mu\nu},~~ \hat G_s = |c S + d|^2 G_s, ~~ \hat \Theta^{(lc)-1}
= \frac{c \chi + d}{|c S + d|^2} \Theta^{(lc)-1},
\ee
where  $G^{(lc)}_{\mu\nu}$,  $G_s$ and $\Theta^{(lc)-1}$ are the
corresponding quantities in $1^{(lc)})$. So this theory looks similar to
 the original theory but it is always strongly coupled. The $SL(2,Z)$
duality here is not useful.
b) If $\chi$ is rational, we
can choose $c \chi + d = 0$. Now $|c \lambda + d| = |c|/g_s =
|c|\a'/(G_s \tilde b)$. $|c S + d|^2 = c^2 /G_s^2$. We have now
\be
\hat G^{(lc)}_{\mu\nu} = \frac{\tilde b}{\a'} \frac{|c|]}{G_s}
G^{(lc)}_{\mu\nu},~~\hat G_s = c^2/G_s,
\ee
and all the noncommutativity parameters vanish. If the original light-like
NCYM is 
strongly coupled, we end up with a physically equivalent weakly coupled 
theory defined on a commutative geometry. Also we have $p \circ p = 0$
since $\Theta^{(lc)\mu\nu}$ vanish. So it appears that
we end up with a light-like OYM. We will comment on this later in this 
subsection.
\item $2^{(lc)})'$ We have also two cases: a) irrational $\chi$ and b)
 rational $\chi$. 

a){\bf Irrational} $\chi$: Now $|c \lambda + d|$ is
fixed since we have fixed $g_s$. So is $|c S + d|^2$. Therefore the
metric $\hat G^{(lc)}_{\mu\nu}$ is essentially the same as the original
metric $G^{(lc)}_{\mu\nu}$. We also have
\be
\hat \Theta^{(lc)-1} = 2\sqrt{2} \pi \tilde b \frac{c \chi + d}{|c S +
d|^2},~~\hat \Theta^{(lc)- 3} = \frac{2\sqrt{2} \pi c \tilde b}{G_s |c S +
d|^2}.
\ee
We again have only the space-space noncommutativity in the light-like 
coordinate system but we double the number of noncommutative pairs. This 
theory is also unitary since $p \circ p \ge 0$. We again end up with a 
light-like NCYM. However, this NCYM is strongly coupled regardless of
 the coupling of the original theory. So the $SL(2,Z)$ duality is not
that useful except for increasing the noncommutative directions.

b){\bf Rational} $\chi$: We can now choose $c \chi + d = 0$. Since
$g_s$ is fixed, we can obtain the corresponding quantities simply by
setting $c\chi + d = 0$ in a). We then have $\hat \Theta^{(lc)-1} = 0$
and $\hat G_s = c^2/G_s$. So we end up with a weakly coupled light-like
NCYM if the original light-like NCYM is strongly coupled. The number of 
noncommutative pairs remain the same but we change from
$\Theta^{(lc)-1}$ to $\hat \Theta^{(lc)-3}$ by the $SL(2,Z)$ duality. 
\item $3)'$ We have two cases: a) irrational $\chi$ and b)
rational $\chi$. \\
a){\bf Irrational} $\chi$: $|c \lambda + d| = |c|
\tilde b /(\a' G_s)$. $|c S + d|^2$ remains fixed. We then have from
\eqn{28} and \eqn{29}
\bea
&&\hat G_{\mu\nu} = \frac{\a'}{\tilde b}~ \frac{G_s}{|c|} |c S + d|^2 
G_{\mu\nu},~~ \hat G_s = |c S + d|^2 G_s,\nnu\\
&& \hat\Theta^{03} = - \frac{2\pi c \tilde b}{\sqrt{g_0} G_s |c S +
d|^2},~~ \hat \Theta^{12} = - \frac{2\pi \tilde b (c \chi + d)}{|c S +
d|^2},~~\hat\Theta^{23} = - \frac{2\pi c \tilde E \tilde b}{G_s |c S + d|^2},
\eea
Since $\a' G^{-1}$ is fixed, we therefore have NCOS. Depending on
the value of $\tilde E$, we can have as many as three independent
noncommutativity parameters. However, this theory is strongly coupled
regardless of the original theory. So it is again not useful. \\
b){\bf Rational} $\chi$:  For this case we can choose $c\chi + d = 0$. 
Now $|c \lambda + d|$ remains the same as the above since
$g_s \to 0$ and $|c S + d|^2 = c^2/G_s^2$ still remains fixed. So
we still have $\hat G_{\mu\nu} \sim \a' G_{\mu\nu}$ which implies that
we still end up with a NCOS. But now $\hat\Theta^{12} = 0$. We have
\be
\hat\Theta^{03} = - \frac{2\pi G_s \tilde b}{c \sqrt{g_0}},~~
\hat\Theta^{23} = - 2\pi G_s \tilde E \tilde b /c.
\ee
We then have 
a weakly coupled NCOS if the original NCYM is strongly coupled. So now
the $SL(2,Z)$ duality is useful.    
\end{itemize}

So far it appears that $1^{(lc)})$ and $2^{(lc)})$ differ from 3) in
that the former are light-like NCYM while the latter is usual NCYM.
Also their $SL(2,Z)$ dualities are quite different. The former give
either light-like NCYM or OYM while the latter gives the usual NCOS.
There is also a big difference regarding the closed string coupling
$g_s$. The former have either $g_s$ blowing up or fixed while the latter
has vanishing $g_s$. In this aspect, these light-like NCYM are similar
to the NCYM
discussed in \cite{lursone} whose S-duality gives an OYM which is not 
well-defined because of the singular metric and infinitely large open
string coupling. We will discuss the $SL(2,Z)$ duality of this kind of 
NCYM in the following subsection.  
There is a difference 
regarding the open string coupling. The $SL(2,Z)$ duality of the
light-like NCYM has a finite (maybe large) open string coupling while
that of  the above mentioned NCYM discussed in \cite{lursone} has
blowing up open string coupling. The reason for this is
that we here consider ${\bf E} \perp {\bf B}$ case and ${\bf E}$ and
${\bf B}$ have no contribution to the quantity $S$. But for ${\bf
E}||{\bf B}$ case, we do have $\tilde E \tilde B$ contribution to the
$S$ as indicated before which blows up in the decoupling limit for the
above mentioned NCYM in \cite{lursone}.

	Up to now we have avoided pointing out the underlying major difference 
between case
$1^{(lc)})$ and $2^{(lc)})$  and case 3). The 
$\det G^{(lc)}_{\mu\nu} \sim \a'^2$ vanishes for the former two cases
while $\det G_{\mu\nu}$ is finite for case 3). In other words, the
light-like NCYM, if they indeed exist,  for the former two cases are
defined on a zero-size spacetime, or singular spacetime, while the
latter is a well-defined usual NCYM. Because of this, at least we have
one component of $\a' G^{\mu\nu}$ nonvanshing (the same is true in the
$SL(2,Z)$ dual). One would say that the underlying theory may not be 
a field theory. One may wonder that the unitarity condition 
obtained from a one-loop analysis in \cite{gomm,ahagm} is sufficient to
show the existence of such light-like NCYM. Further study is needed. 

	One could have non-singular metric by rescaling the light-cone 
coordinate $x^-$. For $1^{(lc)})$, if we rescale $x^- = 
(\tilde b/(\a'\sqrt{g_0g_2})) \tilde x^-$, then we get the open string
metric $G^{(lc)}_{\mu\nu} = {\rm diag}(1/2, 1, -2, 1)$, which is
 non-singular with
respect to the coordinates $(x^+, x^1, \tilde x^-, x^3)$. For
$2^{(lc)})$, if we rescale $x^- = (\tilde b/\a') x^-$, we have, with respect
 to $( x^+, x^1, \tilde x^-, x^3)$,
\be
G^{(lc)}_{\mu\nu} = \left(\begin{array}{cccc}
2&0&-1&0\\
0&1&0&0\\
-1&0&0&0\\
0&0&0&1\end{array}\right),
\lll{48}\ee
which is also  non-singular.
Note that the only noncommutativity parameter with respect to $(x^+, x^1, x^-,
 x^3)$ is $\Theta^{ - 1}$. Actually, we have the two-point function
 $\langle x^- (\tau) x^1 (0)\rangle = (i/2) \Theta^{-1} \epsilon (\tau)$.
The scaling $x^- \sim (1/\a') \tilde x^-$ implies $\langle \tilde x^- x^1
\rangle \sim \a'
 \Theta^{- 1} \epsilon (\tau) \to 0$. Since now $\a' G^{-1} \to 0$
with respect to $( x^+, x^1,\tilde x^-, x^3)$, we therefore end up with
light-like OYM\footnote{We are not sure whether the resulting light-like
 OYM is physically equivalent to the original theory defined on a zero
 size 4-dimensional spacetime because the rescaling of $x^-$ is singular.}
for both 1) and 2).

    Because we end up with OYM for 1) and 2), their $SL(2,Z)$ dualities
still give OYM as discussed at the outset of this section. We will not
give the detail here.

	We now move on to discuss possible NCOS limit and its $SL(2,Z)$
duality. To have decoupling limit for NCOS, we need to keep $\a'
G^{\mu\nu}$ and at least $\Theta^{01}$ fixed when the limit 
$E \to  E_c$ is taken with $E_c$ the critical field
limit. In general, we do not need to send $\a' \to 0$ since the open
string massive modes are not decoupled from its massless modes. However,
it is convenient to choose the $\a' \to 0$ limit since we will study the
$SL(2,Z)$ duality of the resulting NCOS which might be a field theory. 
Now we have a fixed $\a'_{eff}$
for the NCOS which is determined by the noncommutative scale.

    For ${\bf E} \perp {\bf B}$, the critical electric field limit is
$\tilde E^2  \to 1 + \tilde B^2$. From the previous discussion for NCYM, 
we expect that for either fixed $\tilde B$ or infinitely large $\tilde B$ 
as $\a' \to 0$, we have similar complications here. We will discuss
these cases elsewhere.  We here focus on the limit $\tilde B \to 0$ 
along with the above
critical electric field limit as $\a' \to 0$ for NCOS. 
The relation between the
effective open string coupling and the closed string coupling \eqn{24}
implies $\tilde E^2 < 1 + \tilde B^2$, so we should have in general
$\tilde E^2 = 1 + \tilde B^2 - (\a'/\tilde b)^\delta$ with $\delta > 0$
and $\tilde b$ fixed.
With the above discussion, we must also have $\tilde B^2 = (\a'/\tilde
b')^\b$ with $\b > 0$ and $\tilde b'$ fixed. For $\b > \delta$, the
effect of ${\bf B}$ simply drops out and we have purely electric field
effect which has been discussed before \cite{seist,gopmms} and we will not
repeat
this case here. The only other case which gives $G_{\mu\nu} \sim \a'$
is $\b = \delta$. We have two cases: a) $\tilde b > \tilde b'$ and
 b)$\tilde b < \tilde b'$. Let us discuss each in order. a) We now have
the decoupling limit
\bea
&&\a' \to 0,~~ g_0 = \left[\left(\frac{\tilde b}{\tilde b'}\right)^\delta
- 1\right]^{-1} \left(\frac{\a'}{\tilde b}\right)^{1 - \delta},~~
g_1 = \left(\frac{\a'}{\tilde b}\right)^{1 - \delta},\nnu\\
&& g_2 = g_3 = \frac{\a'}{\tilde b},~~g_s = G_s \left(\frac{\tilde
b}{\a'} \right)^{\delta/2},~~ \tilde E^2 = 1 + \left[\left(\frac{\tilde
b}{\tilde b'}\right)^\delta - 1\right] \left(\frac{\a'}{\tilde
b}\right)^\delta,
\lll{52}\eea
with $\tilde B$ given above. We then have
\be
G_{\mu\nu} = \frac{\a'}{\tilde b} \left(\begin{array}{cccc}
1&0&-\left[1 - (\tilde b'/\tilde b)^\delta\right]^{-1/2}&0\\
0&1&0&0\\
-\left[1 - (\tilde b'/\tilde b)^\delta\right]&0&1&0\\
0&0&0&1\end{array}\right),~~
\Theta^{01} = 2\pi
\tilde b \left[\left(\frac{\tilde b}{\tilde b'}\right)^\delta - 1\right]^{1/2}.
\lll{53}\ee
We, therefore, have NCOS with nonvanishing $\Theta^{01}$.
 We now consider case b)\footnote{This case may be equivalent to the one
studied in \cite{chew}.}. The decoupling
limit for this case remains the same except for the scaling for $g_0$
which can be obtained  by the following replacement:
\be
\left(\frac{\tilde b}{\tilde b'}\right)^\delta -
1 \to 
     1- \left(\frac{\tilde b}{\tilde b'}\right)^\delta.
\lll{54}\ee
Now we have 
\be
G_{\mu\nu} = \frac{\a'}{\tilde b}\left(\begin{array}{cccc}
- 1&0& - \left[(\tilde b'/\tilde b)^\delta - 1\right]^{- 1/2}&0\\
0&1&0&0\\
- \left[(\tilde b'/\tilde b)^\delta - 1\right]^{- 1/2}&0&1&0\\
0&0&0&1\end{array}\right),
\ee
 and the nonvanishing
noncommutativity parameter $\Theta^{01}$ which can be obtained from \eqn{53}
by the same replacement as above. We again end up with a NCOS.
 
    We denote the $SL(2,Z)$ duality of the above two cases as $a)'$ and
    $b)'$.
\begin{itemize}
\item $a)'$ We need to consider: 1) irrational $\chi$ and 2) rational
    $\chi$. \\ 1){\bf Irrational} $\chi$: Now since $g_s \to \infty$, we
    have $|c\lambda + d| = |c \chi + d| \neq 0$ and $|c S + d|^2$
    remains fixed. So we have
\bea
&&\hat G_{\mu\nu} = \frac{|c S + d|^2}{|c \chi + d|} G_{\mu\nu},~~ \hat
    G_s = |c S + d|^2 G_s \nnu\\
&&\hat\Theta^{01} = \frac{c\chi + d}{|c S + d|^2} \Theta^{01},~~
\hat\Theta^{23} = - \frac{2\pi \tilde b c}{G_s |c S + d|^2},
\lll{55}\eea
where $G_{\mu\nu}, G_s$ and $\Theta^{01}$ are the open string metric, 
noncommutativity parameter and open string coupling in a) above. Since 
$\a'\hat G^{-1}$ is fixed, so we still have NCOS. This theory is 
strongly coupled
    and again the $SL(2,Z)$ duality is not useful. \\ 2){\bf Rational}
    $\chi$: We can now choose $c \chi + d = 0$. Then we have
$|c \lambda + d| = |c|/g_s = (|c|/G_s) (\a'/\tilde b)^{\delta/2}$ and
$|c S + d|^2 = c^2/G_s^2$ still remains fixed. We then have
\bea
&&\hat G_{\mu\nu} = \frac{|c|}{G_s} \left(\frac{\a'}{\tilde b}
\right)^{ - \delta/2} G_{\mu\nu} ,~~\hat G_s = \frac{c^2}{G_s},\nnu\\
&& \hat \Theta^{23} = - 2\pi \tilde b G_s/c.
\lll{56}\eea
We now have $\a' \hat G^{-1} \to 0$ and nonvanishing noncommutativity
parameter $\hat\Theta^{23}$, therefore we end up with a NCYM. 
This NCYM is weakly coupled if the
    original NCOS is strongly coupled. Therefore the $SL(2,Z)$ duality
    is useful.
\item $b)'$. The discussion for this case is basically the same as in $a)'$
    above and we do not repeat them
    here.
\end{itemize}

\subsection{E $||$ B Case}

	      Unlike the previous one, this case is relatively 
simple since the open string metric is always diagonal and we do not have
the
same complications as we encountered there. Let us begin with the
decoupling limit for NCYM.

For having sensible quantum NCYM, we need to keep the open string
metric, the open string coupling and at least one space-space
noncommutativity parameter fixed as $\a' \to 0$. For simplicity, we choose
$G_{\mu\nu} = \eta_{\mu\nu} = (-1, 1, 1, 1)$. From \eqn{34}, we have
$\tilde E^2 \le 1$. We then have the
following decoupling limit:
\be
\a' \to 0, ~~\tilde B = \frac{\tilde b}{\a'},~~
g_2 = \left(\frac{\tilde b}{\a'}\right)^2,~~g_1 (1 - \tilde E^2) = 1,~~
g_s = \frac{G_s}{\sqrt{1 - \tilde E^2}}~\frac{\a'}{\tilde b}.
\lll{56a}\ee
The only nonvanishing noncommutativity parameter is
\be
\Theta^{23} = - 2\pi \tilde b.
\lll{57}\ee
In the above, we have not specified how $\tilde E$ scales. It appears
that the resulting NCYM does not require this as long as $\tilde E^2 \le
1$. However, the scaling behavior of this parameter has great impact
on its $SL(2,Z)$ dual description. This dual description may have a
small coupling, therefore a good one,
in the case when the open string coupling $G_s$ is large.
For this purpose, let us consider the following three cases which
correspond to those studied in \cite{lursone}:
\begin{itemize}
\item a) $\tilde E$ is fixed but it equals neither 0 nor unity\footnote{
$\tilde E = 0$ corresponds to zero electric field which is not our
interest here. $\tilde E = 1$ gives a singular open string metric and
the NCYM is no longer 1 + 3 dimensional which is not our interest here,
either. So we exclude these two cases here.}.
\item b) $\tilde E = 1 - (\a'/\tilde b')^\delta/2$ with $\delta > 0$.
\item c) $\tilde E = (\a'/\tilde b')^\b$, with $\b > 0$. 
\end{itemize}
We would like to point out that the electric field in b) becomes
critical but does not have effect on NCYM.
  
Let us study each of the above in the $SL(2,Z)$ dual description.

\noindent 
{\bf Case a)}: Using the decoupling limit in \eqn{56a}, we have $|c \lambda + d|
=
     c /g_s = c \tilde b (1 - \tilde E^2)^{1/2}/ (\a' G_s)$ and 
$|c S + d| = c \tilde B \tilde E /G_s = c \tilde b \tilde E /(\a' G_s)$.
Using these we have 
\be
\hat G_{\mu\nu} \sim \eta_{\mu\nu} /\a',~~\hat\Theta^{01} \sim \a'^2,~~
\hat\Theta^{23}\sim \a'~~\hat G_s \sim 1/\a'^2.
\lll{58}\ee   
Since $\a' \hat G^{\mu\nu} \sim \eta^{\mu\nu} \a'^2 \to 0$,  we still
have a field theory but defined in a commutative geometry. However, this
theory is bad since it has an infinitely large open string coupling 
and a singular metric. Even if we rescale the coordinates to have a finite
metric but we cannot change the open string coupling. So we cannot turn
this theory to a well-defined one. We here reach the same conclusion as
in \cite{lursone} regardless of the fact that $\chi$ is rational or not.

\noindent
{\bf Case b)}: This case is not much different from case a). Even though the
scaling of the open string metric  depends on whether $\chi$
is rational or not, it always blows up as $\a' \to 0$. So we still end
up with a field theory which is not well-defined since the open string
coupling blows up in the same way as in case a). The noncommutativity
parameters scale as
\be
\hat\Theta^{01}\sim \a'^{2 + \delta},~~\hat\Theta^{23} \sim \a'.
\lll{59}\ee

\noindent
{\bf Case c)}: From our experience in \cite{lursone} on S-duality, we expect
that this is the case for which we expect to have NCOS. 
 We now have
$g_s = \a' G_s /\tilde b$ and $|c \lambda + d| = c/g_s = c \tilde b
/(\a' G_s)$. We have three sub-cases to consider: $0 < \b < 1$, $\b = 1$
and $\b >
1$. For $0 < \b < 1$, we reach the same conclusion as in case a) and b)
above, i.e., we end up with a field theory which is not well-defined 
because of the infinitely large open string coupling. This subcase has   
also been studied in \cite{lursone} on S-dual rather than on $SL(2,Z)$
dual. The conclusion remains the same and we will skip the details.
We now focus on $\b = 1$ and $\b > 1$ subcases. For $\b = 1$, $|c
S + d|^2 = [(c \chi + d) + \tilde b /(\tilde b' G_s)]^2 + c^2 /G_s^2$
is fixed and we have 
\bea
&&\hat G_{\mu\nu} = \frac{\a'}{\tilde b} \frac{G_s} {|c|} |c S + d|^2  
\eta_{\mu\nu},~~\hat G_s = G_s |c S + d|^2,\nnu\\
&&\hat\Theta^{01} = -\frac{2\pi \tilde b}{G_s |c S +
d|^2},~~\hat\Theta^{23} = - \frac{2\pi \tilde b}{|c S + d|^2}\left[
(d + c \chi) + \frac{c\tilde b}{G_s \tilde b'}\right].
\lll{60}\eea
The above implies that $\a' \hat G^{\mu\nu}$ is fixed. We therefore have 
NCOS rather than NCYM. In other words, the $SL(2,Z)$ dual of NCYM for
$\b = 1$ gives a NCOS whether  $\chi$ is rational or not. This is due to
$g_s \to 0$. However,
whether $\chi$ is rational or not is important in determining the 
usefulness of the $SL(2,Z)$ duality. Our primary purpose 
 is to find a weakly coupled theory by $SL(2,Z)$ duality when the
open string coupling for NCYM is large. When $\chi$ is irrational, we
map a strongly coupled theory (NCYM) to another strongly coupled
theory (NCOS) by $SL(2,Z)$ duality which can be examined from
 the relation between two open string couplings given in \eqn{60}.
So $SL(2,Z)$ duality is not particularly useful in this case. However,
when $\chi$ is rational, we can always choose $c \chi + d = 0$ through
$SL(2,Z)$ duality. Then we can map a strongly coupled theory (NCYM) to a 
physically equivalent and weakly coupled theory (NCOS). So only for 
rational $\chi$, the S-duality is useful.

For $\b > 1$, we continue to have $|c\lambda + d| = |c|/g_s = 
|c| \tilde b/(\a'G_s)$ but now $|c S + d|^2 = (c\chi + d)^2 + c^2/G_s^2$.
We then have
\bea
&&\hat G_{\mu\nu} = \frac{\a'}{\tilde b}~\frac{G_s}{|c|} |c S + d|^2 
\eta_{\mu\nu}, ~~\hat G_s = G_s |c S + d|^2,\nnu\\
&&\hat\Theta^{01} = - \frac{2\pi \tilde b}{G_s |c S + d|^2},~~  
\hat\Theta^{23} = - \frac{2\pi (c \chi + d)}{|c S + d|^2}.
\lll{62}\eea
We have again NCOS since $\a' \hat G^{\mu\nu} \sim \eta^{\mu\nu}$. Only for 
rational $\chi$, a strongly coupled NCYM can be mapped to a weakly
coupled NCOS by $SL(2, Z)$ duality since we can  choose
$c \chi + d = 0$. Once such a choice is made, 
we have $\hat \Theta^{23} = 0$.
This case is not different from the one with $\tilde E = 0$.
However, when $\chi$ is irrational, we end up not only with a strongly
coupled theory  but also
with nonvanishing $\hat\Theta^{23}$ even if we start with $\tilde E =
0$.

Let us now discuss the $SL(2,Z)$ duality of NCOS. To have NCOS, we need
$\a' G^{\mu\nu}$ and at least $\Theta^{01}$ to be fixed when the
critical electric field limit $\tilde E \to 1$ is taken. Unlike in the 
field theory limit, we do not need to take $\a' \to 0$ since we do not 
require the open string massive modes to decouple from its massless
ones. Our purpose here is to study the $SL(2,Z)$ dual of NCOS which
might be a NCYM. For this reason, it is convenient to set $\a' \to 0$
for NCOS such that we can easily discuss its $SL(2,Z)$ dual which has
the possibility of NCYM. In doing so, the effective open string
$\a'_{\rm eff}$ for the NCOS is still fixed and is determined by the 
noncommutativity parameter or scale. Since we require $\a' G^{\mu\nu}$ to
be fixed as $\a' \to 0$, so we can set for simplicity
\be
G_{\mu\nu} = \frac{\a'}{\tilde b} \eta_{\mu\nu}.
\lll{63}\ee
From eqs.\eqn{31}-\eqn{34}, we have the following decoupling limit:
\bea
&&\a' \to 0,~~\tilde E = 1 - \frac{1}{2}\left(\frac{\a'}{\tilde b'}
\right)^\delta,~~g_1 = \frac{\tilde b'}{\tilde b}\left(
\frac{\a'}{\tilde b'}\right)^{1 - \delta},\nnu\\
&&g_2 (1 + \tilde B^2) = 
\frac{a'}{\tilde b},~~g_s = \left(\frac{\tilde
b'}{\a'}\right)^{\delta/2} \frac{G_s}{(1 + \tilde B^2)^{1/2}},
\lll{64}\eea
with $\delta > 0$.
From the above and \eqn{33}, we have
\be
\Theta^{01} = 2\pi \tilde b,~~\Theta^{23} = - 2\pi\tilde b \tilde B.
\lll{65}\ee
In the above, we have not yet specified how $\tilde B$ scales. In
general we can set $\tilde B = h (\a'/\tilde b')^\b$ with $h$ fixed and
$\b \ge 0$. For $\b = 0$,  $\Theta^{23}$ is fixed while it vanishes for 
$\b > 0$. Using this decoupling limit, we try to find the underlying
theory after $SL(2,Z)$ duality. Whether $\chi$ is rational or not
is 
crucial for the conclusion. So we discuss them separately in the
following.

\noindent

{\bf Irrational} $\chi$: In this case we have $ c \chi + d \neq 0$.
Now $|c \lambda + d| = |c \chi + d|$ 
since $g_s \to \infty$. For $\b = 0$, 
$|c S + d|^2 = [c(\chi + h/G_s) + d]^2 + c^2/G_s^2$. Whereas for
$\b > 0$, $|c S + d| = (c \chi + d)^2 + c^2/G_s^2$. So for $\b \ge 0$,
$|c S + d|$ is fixed. From \eqn{37}, \eqn{38} and \eqn{39}, we have 
\bea
&&\hat G_{\mu\nu} = \frac{\a'}{\tilde b} \frac{|c S + d|^2}{|c \chi + d|}
\eta_{\mu\nu},~~ \hat G_s = |c S + d| G_s,\nnu\\
&& \hat \Theta^{01} = \frac{(c \chi + d)}{|c S + d|^2} \Theta^{01},~~
\hat \Theta^{23} = \frac{(c \chi + d)\Theta^{23} - 2 \pi c \tilde b (1 +
\tilde B^2)/G_s}{|c S + d|^2}.
\lll{66}\eea
The scaling of the metric $\hat G_{\mu\nu}$ tells that we end up
actually with a NCOS rather than a NCYM for irrational $\chi$. This has 
been given first in \cite{russj}. Notice
that we now have nonvanishing $\hat \Theta^{23}$ even if we begin with
$\Theta^{23} = 0$. However, we map a strongly coupled NCOS to another
strongly coupled NCOS. Therefore, for irrational $\chi$, the $SL(2,Z)$
duality is not that useful. The interesting point in this case is that we
can use it 
to reduce or to increase the space-space noncommutative directions 
(since we can get a vanishing $\Theta^{23}$ from a nonvanishing 
$\hat \Theta^{23}$ or vice-versa).

\noindent  
 
{\bf Rational} $\chi$: Now we can always choose $c \chi + d = 0$.
Then $|c\lambda + d| = c/g_s = (\a'/\tilde b')^{\delta/2} c (1 +
\tilde B^2)^{1/2}/G_s$. Again $|c S + d|^2$ remains fixed. So we have
\be
\hat G_{\mu\nu} \sim \a'^{1 - \delta/2}\eta_{\mu\nu},~~ \hat \Theta^{01} = 0,~~
\hat\Theta^{23} = - \frac{2\pi c \tilde b (1 + \tilde B^2)}{G_s |c S +
d|},
\lll{67}\ee
and the open string coupling $\hat G_s = c^2 (1 + h^2)/G_s$ for $\b = 0$
and $\hat G_s = c^2/G_s$ for $\b > 0$.

    Since $\a' \hat G^{-1} \sim \a'^{\delta/2} \to 0$, we therefore end
 up with a NCYM with noncommutative space-space directions. This has
 also been studied in \cite{russj}. So now a
 strongly coupled NCOS is physically equivalent to a weakly coupled 
NCYM. In this case, the $SL(2,Z)$ is really useful and now a $SL(2, Z)$
is not much different from a simple S-duality as studied in \cite{lursone}.
A space-time noncommutativity is also transformed to a space-space one.
   
\section{Conclusion}

	To conclude, we have discussed in this paper  various
decoupling limits for noncommutative open
	string/Yang-Mills theory in four-dimensions and their $SL(2,Z)$ 
duality for both ${\bf E}\perp{\bf B}$ and ${\bf E} || {\bf B}$
	cases. Since $SL(2,Z)$ is a non-perturbative quantum symmetry of
	type IIB string theory, we often use this symmetry to find a physically
	equivalent and yet weakly coupled theory if the original theory
	is strongly
	coupled. However, our study indicates that if the RR scalar in
	one theory is irrational, the $SL(2,Z)$ does not help much
	and the $SL(2,Z)$ dual is always strongly coupled. So when we
	say that we can use S-duality or in general $SL(2,Z)$ duality to 
transform a strongly coupled theory to a weakly coupled one, one must
	understand that this can be done only for rational $\chi$. Since
$\chi$ is determined by the underlying (most likely non-perturbative)
	vacuum, whether $\chi$ is rational or not is a rather
	non-trivial question. We cannot answer this until we
	understand the non-perturbative type IIB theory completely.

We also find that $SL(2,Z)$ symmetry can be used to increase or decrease
the number of noncommutative directions but it seems that we cannot
turn an OYM to a NCYM/NCOS through this symmetry. We also find that 
the interplay of electric and magnetic fields are important in
controlling the number of noncommutative directions. We show that the
$SL(2,Z)$ duality of NCYM can be an ordinary theory which is not 
well-defined or
a NCOS regardless of whether $\chi$ is rational or not. But only for
rational $\chi$, the NCOS can be weakly coupled if the original NCYM is 
strongly coupled. Also when the original NCOS is strongly coupled the 
$SL(2,Z)$ duality is either another strongly
coupled NCOS if $\chi$
is irrational or a weakly coupled NCYM if $\chi$ is rational.
Some of the critical electric field
limit for ${\bf E}\perp {\bf B}$ are particularly interesting. Whether
we have decoupled light-like NCYM or light-like NCOS or other kinds of
OS or light-like OYM is still not clear. Further study is needed.

\bigskip
\center{\bf ACKNOWLEDGMENTS}

JXLU acknowledges the support of U. S. Department of Energy.

\end{document}